\documentclass[aps,prc,twocolumn,lengthcheck,showpacs,showkeys,amsfonts,amssymb,amsmath,nofootinbib,a4paper,oneside]{revtex4-1}
\usepackage{bm,amsmath,epsf,epsfig,latexsym}
\usepackage{graphics,psfrag}
\usepackage{hyperref}
\hypersetup{colorlinks=true}
\hypersetup{urlcolor=black}
\hypersetup{linkcolor=blue}
\hypersetup{citecolor=blue}
\newcommand{\dd}{\mathrm{d}}

\newcommand{\vp}{{\mathbf{p}}}

\newcommand{\bg}{\begin{align}}
\newcommand{\eeg}{\end{align}}
\newcommand{\be}{\begin{equation}}
\newcommand{\ee}{\end{equation}}
\newcommand{\ba}{\begin{eqnarray}}
\newcommand{\ea}{\end{eqnarray}}

\newcommand{\nn}{\nonumber}

\newcommand{\la}{\langle}
\newcommand{\ra}{\rangle}
\newcommand{\si}{\sigma}
\newcommand{\vz}{\mathbf{z}}

\newcommand{\jls}{J \ell S}
\renewcommand{\ge}{\geqslant}

\renewcommand{\geq}{\geqslant}
\renewcommand{\leq}{\leqslant}
\newcommand{\infff}{{\bm{\infty}}}

\usepackage[T1]{fontenc}
\usepackage{textcomp}

\usepackage{mathptmx}

\usepackage{fancyhdr}

\allowdisplaybreaks

\usepackage{titlesec}
\titlespacing{\section}{0pt}{20pt}{5pt}
\titlespacing{\subsection}{0pt}{20pt}{5pt}
\titleformat{\section}[hang]
{\filcenter}
{\bfseries\thesection.}
{5pt}
{\filcenter\bfseries\MakeUppercase}
\begin{document}

\frenchspacing
\parskip 0cm

\title{Nucleon-nucleon interactions from dispersion relations: Elastic partial waves}
\author{M.~Albaladejo}
\email{albaladejo@um.es}
\author{J.~A.~Oller}
\email{oller@um.es}
\affiliation{Departamento de F\'{\i}sica, Universidad de Murcia, E-30071, Murcia, Spain.}

\begin{abstract}
  We consider nucleon-nucleon ($NN$) interactions from chiral effective field theory. In this work we restrict ourselves to the elastic $NN$ scattering. We apply the $N/D$ method to calculate the $NN$ partial waves taking as input the one-pion exchange discontinuity along the left-hand cut. This discontinuity is amenable to a chiral power counting as discussed by Lacour, Oller and Mei{\ss}ner [Ann. Phys. (NY) {\bf 326}, 241 (2011)], with one-pion exchange as its leading order contribution. The resulting linear integral equation for a partial wave with orbital angular momentum $\ell\geq 2$ is  solved in the presence of $ \ell-1$ constraints, so as to guarantee the right behavior of the $D$ and higher partial waves near threshold. The calculated $NN$ partial waves are based on dispersion relations and are independent of regulator. This method can also be applied to higher orders in the calculation of the discontinuity along the left-hand cut and extended to triplet coupled partial waves.

\end{abstract}

\maketitle

\section{Introduction}
\label{sec:intro}

The nucleon-nucleon ($NN$) interaction is a basic process whose understanding is necessary for
 the study of nuclear structure, nuclear reactions, nuclear matter, neutron stars, etc. \cite{machleidt_rep,hammer,long,let1,lacour1}.  
 Since the early 1990s \cite{w1,w2,w3} the low energy effective field theory (EFT) of QCD, chiral perturbation theory (ChPT), 
has been applied to  $NN$ scattering in a large number of studies \cite{w1,w2,w3,ordo94,kolck94,kolck99,kaiser97,kaiser98,epelbaum1,bedaque,entem,mach,epen3lo}. A sophisticated stage has been reached where the $NN$ potential is calculated in ChPT up to N$^3$LO \cite{entem,epen3lo}.

  However, as the $NN$ interaction is nonperturbative, the chiral $NN$ potential must be iterated. It was proposed by Weinberg in his seminal papers \cite{w1,w2} to solve a Lippmann-Schwinger equation. Since the chiral potential is singular at the origin a regularization method, typically a three-momentum cut-off $\Lambda$ \cite{ordo94,entem,epen3lo}, should be introduced for solving the Lippmann-Schwinger equation. Despite the great phenomenological success achieved by the $NN$ chiral potentials in describing $NN$ scattering data  \cite{entem,epen3lo}, a remnant  cut-off dependence is left. It was shown in the literature \cite{nogga,pavon06,pavon06b,entem08,kswa} that the chiral counterterms introduced in the $NN$ potential following naive dimensional analyses are not enough to renormalize the resulting $NN$ scattering amplitude. In Ref.~\cite{nogga} one  counterterm is promoted from higher to lower orders in the partial waves $^3P_0$, $^3P_2$ and $^3D_2$ due to the attractive $1/r^3$ tensor force from one-pion exchange (OPE). As a consequence, stable results, independent of cut-off in the limit $\Lambda\to \infff$, are achieved when the LO OPE potential is employed. Partial wave amplitudes with larger orbital angular momentum $\ell$, $\ell\geq 3$, can be  calculated in Born approximation with sufficient accuracy \cite{birse,kaiser97}. Then they do not pose any problem for renormalization, making use of standard perturbative techniques.  It is also argued in the same Ref.~\cite{nogga} that higher orders terms in the chiral $NN$ potential could be treated perturbatively. Reference \cite{nogga} was extended along these lines to subleading two-pion exchange (TPE) in Ref.~\cite{pav11}. The promotion of higher orders to lower ones due to nonperturbative renormalizability is studied in detail in Ref.~\cite{birse} by making use of the regularization group equations. (See also Refs.~\cite{arriola06,pavon06,pavon06b} for a coordinate space renormalization by imposing appropriate boundary conditions.) On the other hand,  Refs.~\cite{epege,epemeiss06}, following the philosophy of Refs.~\cite{lepa1,lepa2}, stress that the cut-off $\Lambda$ should not be taken beyond the breakdown scale of the EFT, typically around 1~GeV. It is argued that if this is done the power counting associated with the chiral EFT is lost. 

  We employ here the $N/D$ method \cite{chew} for studying $NN$ interactions in the elastic case. A linear integral equation then results for determining the $NN$  partial waves. The input is given by the discontinuity of the partial wave along the left-hand cut (LHC), which is due to multipion exchanges, the lightest one being OPE. The well-known behavior of a partial wave near threshold, that vanishes like $|\vp|^{2\ell}$, with $|\vp|$ the center of mass (c.m.) three-momentum, is not automatically fulfilled in the $N/D$ method for $\ell\geq 2$  \cite{barton,morav,bala}. Then, the $N/D$ method must be solved in the presence of $\ell-1$ constraints. These are satisfied by introducing $\ell-1$ Castillejo-Dalitz-Dyson (CDD) poles \cite{cdd}, as we discuss below. The extension of the work to the inelastic partial waves will be considered separately \cite{albaprepare}, due to the intrinsic specific difficulties that occur in that case. The latter do not appear for the elastic partial waves which then allows us to focus in those aspects that directly concern the application of the $N/D$ method to $NN$ scattering.

After this Introduction we discuss the $N/D$ method for calculating the $NN$ elastic partial waves in Sec.~\ref{sec:method2}. Special attention is paid to derive the constraints needed to meet the threshold behavior for a partial wave with $\ell\geq 2$. The inclusion of CDD poles for satisfying these constraints is a novelty in the literature. The results are discussed in Sec.~\ref{sec:results} and conclusions are given in Sec.~\ref{sec:conclusions}. The Appendix discusses the numerical method employed to solve the integral equation.

\section{Application of the $N/D$ Method to Elastic $NN$ Partial Waves}\label{sec:method2}\subsection{$NN$ partial waves cuts}We consider the $NN$ interaction process
\begin{align*}  
N(\vp_1;\sigma_1\alpha_1) N(\vp_2;\sigma_2\alpha_2)\to N(\vp'_1;\sigma_1'\alpha_1') N(\vp'_2;\sigma_2'\alpha_2')
\end{align*}
 whose scattering amplitude in the c.m. frame is indicated by 
\begin{align*}
\la \vp',\si'_1 \alpha'_1 \si'_2 \alpha'_2 |T_d| |\vp|\hat{\vz}, \si_1\alpha_1
\si_2\alpha_2\ra~.
\end{align*}
 Here the initial momentum is $\vp=|\vp|\hat{\vz}$, taken along the $z$ axis, and the final one is $\vp'$. Its decomposition in partial waves is discussed in Appendix A of Ref.~\cite{long}, to which we refer for further details. We denote a $NN$ partial wave by $T_{JI S}(\ell',\ell;|\vp|^ 2)$, being $\ell'$ the final orbital angular momentum and $\ell$ the initial one. The labels $J$, $S$, and $I$ stand for the total angular momentum, spin and isospin of the reaction, respectively:
\begin{align}
T&_{J I S}(\ell',\ell;|\vp|^ 2) \nn\\
&=\frac{Y_\ell^0(\hat{\vz})}{2J+1}\sum (\si_1'\si_2's'_3|s_1 s_2 S)(\si_1\si_2 s_3|s_1 s_2 S)(0 s_3 s_3|\ell S J) \nn\\
& \times (m' s'_3 s_3|\ell' S J)(\alpha'_1\alpha'_2 i_3|\tau_1\tau_2 I)(\alpha_1\alpha_2 i_3|\tau_1\tau_2 I) \nn\\
& \times \int d\hat{\vp'}\,\la \vp',\si'_1 \alpha'_1 \si'_2 \alpha'_2 |T_d| |\vp|\hat{\vz}, \si_1\alpha_1 \si_2\alpha_2\ra Y_{\ell'}^{m'}(\vp')^*~.
\label{pw.exp.def}
\end{align} 
In this equation, the Clebsch-Gordan coefficients for the couplings of two angular momentum $j_1$, $j_2$ to $j_3$ are indicated by $(m_1 m_2 m_3|j_1 j_2 j_3)$, with $m_1$, $m_2$ and $m_3$ the corresponding third components.

\begin{figure}[t]
\includegraphics[width=6cm,keepaspectratio]{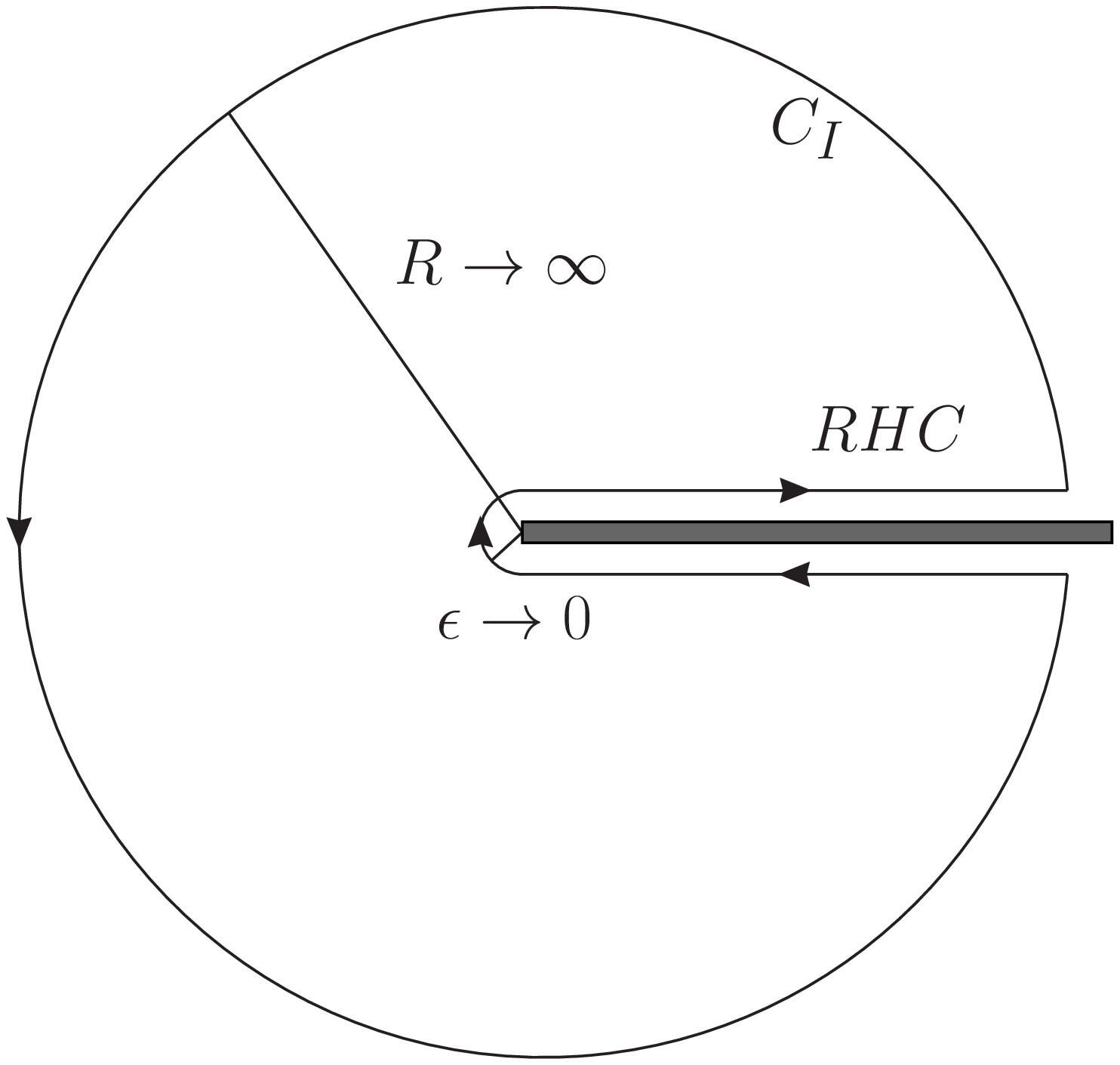}\\
\includegraphics[width=6cm,keepaspectratio]{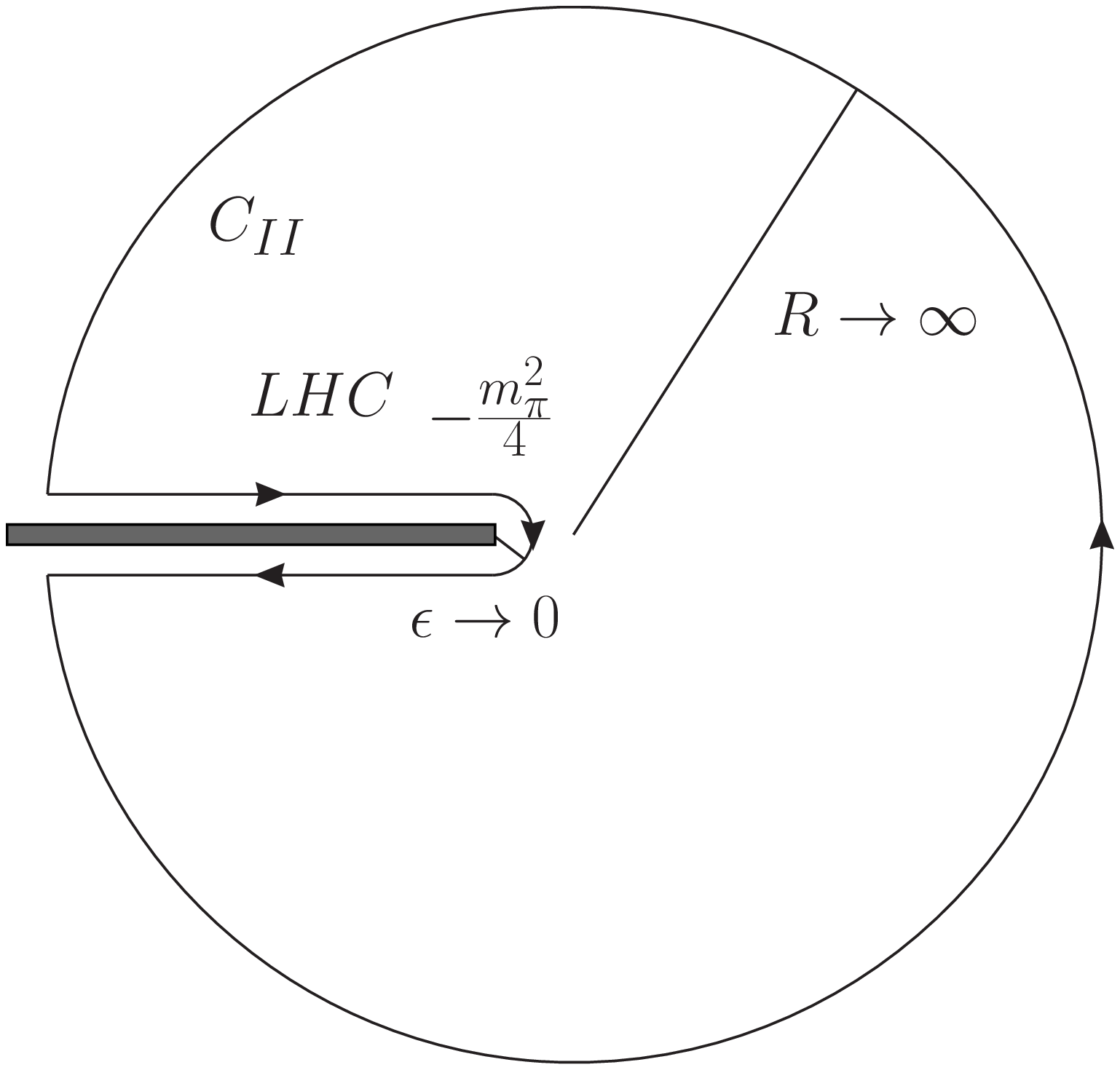}
\caption{The thick lines correspond to the RHC and LHC, in order from top to bottom. In the same figure the integration contours $C_I$ and $C_{II}$ for evaluating $D_{J\ell S}(A)$ and $N_{J\ell S}(A)$, 
respectively, are shown. One has to take the limit $\epsilon\to 0^+$\label{fig:merge}}
\end{figure}

A $NN$ partial wave amplitude  has two cuts \cite{spearman}, the right-hand cut (RHC) for $0< \vp^2 <\infff$, due to unitarity, and the LHC for $-\infff<\vp^2<L$ with $L=-m_\pi^2/4$, because of crossed channel dynamics. Both cuts are depicted in Fig.~\ref{fig:merge}. The upper limit for the latter  is given by OPE, as the pion is the lightest particle that can be exchanged in the $t$ and $u$ channels. Because of unitarity a partial wave amplitude satisfies in the c.m. frame, above the elastic threshold and below pion production, the relation
\be
\hbox{ Im}T_{J I S}(\ell',\ell;|\vp|^2)^{-1}=-\frac{m |\vp|}{4
\pi}\delta_{\ell' \ell}~,
\label{unitarity}
\ee
with $m$ the mass of the nucleon. In our normalization, the $S$-matrix is given by
\begin{equation*}
S_{JIS} = \mathbb{I} + i \frac{m \lvert \vp \rvert}{2\pi} T_{JIS}~\text{.}
\end{equation*}

As shown in Ref.~\cite{long} one can calculate perturbatively in ChPT $\text{Im}T_{J I S}$ along the LHC, since this imaginary part is due to multipion exchanges. The infrared enhancements associated with the RHC are absent in the discontinuity along the LHC because, according to Cutkosky's theorem \cite{landau,cutkosky}, it implies to put on-shell pionic lines. Within  loops the pion poles are picked up, making that the energy along nucleon propagators now is of ${\cal O}(p)$, instead of a nucleon kinetic energy. In this way, the order of the diagram rises compared to that of the reducible parts and it becomes a perturbation. At LO, according to the counting developed in Refs.~\cite{let1,long} (that for two-nucleon irreducible diagrams coincides with the standard chiral counting \cite{w1,w2,w3}), $\text{Im}T_{J I S}$ along the LHC is due to OPE.

\subsection{$S$ and $P$ waves ($\ell = 0,1$)}\label{subsec:sp_waves}
In the following we take the elastic case  for which $\ell'=\ell=J$ (except for the $^3P_0$.) The $N/D$ method \cite{chew} rests on the separation between the RHC and LHC. In this way, a partial wave $T_{J\ell S}(A)$ is written as\footnote{We replace the subscript $I$ by  $\ell$ when denoting a partial wave. The former can be deduced from $\ell$ and $S$.}
\begin{align}
T_{J\ell S}(A)=\frac{N_{J\ell S}(A)}{D_{J\ell S}(A)}~.
\label{nd.def}
\end{align} 
 The function  $N_{J\ell S}(A)$ has only LHC while $D_{J\ell S}(A)$ has only RHC. Taking into account elastic unitarity, Eq.~\eqref{unitarity}, one can write
 \begin{align}
 \text{Im}D_{J\ell S}(A)&=-N_{J\ell S}(A)\frac{m\sqrt{A}}{4\pi}~~,~~A>0~.
\end{align}
Along the LHC one also has from Eq.~\eqref{nd.def} 
\begin{align}
 \text{Im}N_{J\ell S}(A)&=D_{J\ell S}\,(A)\text{ Im}T_{J\ell S}(A)~~,~~A<-m_\pi^2/4~.
 \end{align} 
 We first write down a dispersion relation (DR) for $D_{J\ell S}(A)$ and $N_{J\ell S}(A)$ taking as integration contours $C_I$ and $C_{II}$ in Fig.~\ref{fig:merge}, respectively. The integration along the circle at infinity 
vanishes, if necessary, by taking the sufficient number of subtractions.
 For the case of a once-subtracted DR the following expressions result  
 \begin{align}
 \label{dd}
D_{J\ell S}(A)&=1-\frac{A-D}{\pi}\int_0^{+\infff} \!\!\!\!\!\!\!\! \dd q^2\frac{\rho(q^2)N_{J\ell S}(q^2)}{(q^2-A)(q^2-D)}
~,\\
\label{nn}
N_{J\ell S}(A)&=N_{J\ell S}(D)+\frac{A-D}{\pi}\int_{-\infff}^{L} \!\!\!\!\!\!\!\! \dd k^2
\frac{\Delta_{J\ell S}(k^2)D_{J\ell S}(k^2)}{(k^2-A)(k^2-D)}
~.
 \end{align}
Here we have indicated by $\rho(A)=m \sqrt{A}/4\pi$ for $A>0$ and $\Delta_{J\ell S}(A)=\text{Im}T_{J\ell S}(A)$ for $A<L=-m_\pi^2/4$. For physical scattering $A\to A+i\epsilon$~. The subtraction constant in $D_{J\ell S}(A)$ has been fixed to 1 because, according to Eq.~\eqref{nd.def}, only the ratio between $N_{J\ell S}(A)$ and $D_{J\ell S}(A)$ matters for determining $T_{J\ell S}(A)$. In this way,  one has the freedom to fix the value of $D_{J\ell S}(A)$ at one point. 

Asymptotically, for $\vp^2\to -\infff$, OPE tends to constant, so that, according to the Sugawara and Kanazawa theorem \cite{barton,suga} one subtraction is necessary for the DR of $N_{J\ell S}(A)$, even though $\Delta_{J\ell S}(A)\to 1/A$ in the case of OPE. On general grounds, a partial wave amplitude is bound for $A\to +\infff$ by a constant because of unitarity and the same theorem then requires that at least one subtraction is necessary.

An integral equation for $D_{J\ell S}(A)$ results by inserting the DR for $N_{J\ell S}(A)$, Eq.~\eqref{nn}, into Eq.~\eqref{dd}:
\begin{align}
& D_{J\ell S}(A)=1-N_{J\ell S}(D)\frac{A-D}{\pi}\int_0^{+\infff}\!\!\!\!\!\!\!\!\dd q^2\frac{\rho(q^2)}{(q^2-A)(q^2-D)}\nn\\
&-\frac{A-D}{\pi^2}\int_0^{+\infff}\!\!\!\!\!\!\!\!\dd q^2\frac{\rho(q^2)}{q^2-A}\int_{-\infff}^{L}\!\!\!\!\!\!\!\!\dd k^2\frac{\Delta_{J\ell S}(k^2)D_{J\ell S}(k^2)}{(k^2-q^2-i\epsilon)(k^2-D)}~.
\label{d.1}
\end{align}
We now introduce the function $g(A,C)$ defined as
  \begin{align}
g(A,C)=\frac{1}{\pi}\int_0^{+\infff}\!\!\!\!\!\!\!\!\dd q^2\frac{\rho(q^2)}{(q^2-A)(q^2-C)}~.
 \label{g.def}
 \end{align}
In terms of this function Eq.~\eqref{d.1} can be written as
\begin{align}
&D_{J\ell S}(A)=1- (A-D) N_{J\ell S}(D) g(A,D) \nn\\
& +\frac{A-D}{\pi}\int_{-\infff}^{L}\!\!\!\!\!\!\!\!\dd k^2\frac{\Delta_{J\ell S}(k^2)D_{J\ell S}(k^2)}{k^2-D}g(A,k^2)~.
\label{dis.rel.d}
\end{align}
This a linear integral equation for $D_{J\ell S}(A)$. Its linearity  is an important property because it allows one to take more subtractions and still being amenable for an iterative solution. We take as a convenient subtraction point $D=0$. In the case of the $S$ waves, this choice relates the subtraction constant, $N_{J\ell S}(0)$, to the corresponding scattering length, $a_s$, by
\begin{equation}
N_{J\ell S}(0) = - \frac{4\pi a_s}{m}~.
\end{equation}
The only elastic $S$ wave is the $^1 S_0$, and the scattering length for this wave is $a_s = -23.758 \pm 0.04\ \textrm{fm}$. For this partial wave, we can write down the following integral equation:
\begin{align}
D_{\jls}(A) & = 1 - A N_{\jls}(0)g(A,0) + \nn \\ & + \frac{A}{\pi} \int_{-\infff}^{L}\!\!\!\!\!\!\!\!\dd k^2\frac{\Delta_{J\ell S}(k^2)D_{J\ell S}(k^2)}{k^2}g(A,k^2)~.\label{dd_finalform}
\end{align}

This is an integral equation for $D_{\jls}(A)$ with $A$ on the LHC, and it can be solved with the method exposed in the Appendix. Once $D_{J\ell S}(A)$ is solved from Eq.~\eqref{dd_finalform}, $N_{J\ell S}(A)$ is determined by inserting the former into Eq.~\eqref{nn}, which now reads ($D=0$)
\begin{align}\label{nn_finalform}
N_{J\ell S}(A)&=N_{J\ell S}(0)+\frac{A}{\pi}\int_{-\infff}^{L} \!\!\!\!\!\!\!\! \dd k^2
\frac{\Delta_{J\ell S}(k^2)D_{J\ell S}(k^2)}{k^2(k^2-A)}
~.
\end{align}

For a $P$ wave, the same equations hold with  $N_{\jls}(0) = 0$ because in this case the amplitude vanishes at threshold as $|\vp|^2$, and $D_{J\ell S}(0)=1$.

\subsection{Higher waves ($\ell \ge 2$)}
Equations~\eqref{dd_finalform} and \eqref{nn_finalform}, because of the full implementation of rescattering in Eq.~\eqref{dis.rel.d}, do not guarantee that the resulting partial wave amplitude behaves as $A^{\ell}$ for $A\to 0$,  with $\ell\geq 2$. At LO $\Delta_{J\ell S}\equiv \Delta_{J\ell S}^{1\pi}$ gives rise to OPE through the dispersive integral
\begin{align}
T_{J\ell S}^{1\pi}(A)=T_{J\ell S}^{1\pi}(0)+\frac{A}{\pi}\int_{-\infff}^L\!\!\!\!\!\!\!\!\dd k^2 \frac{\Delta_{J\ell S}^{1\pi}(k^2)}{k^2(k^2-A)}~,
\label{1pi}
\end{align}
with $T_{J\ell S}^{1\pi}(0)$ a subtraction constant. As discussed above, since the OPE amplitude \cite{long} tends to constant for $A\to \infff$, the Sugawara and Kanazawa theorem requires that one subtraction is needed. The fact that for $\ell >0$ a partial wave vanishes  as $A^\ell$ for $A\to 0$ makes that $T_{J\ell S}^{1\pi}(0)=0$ when  $\ell>0$. This threshold behavior also implies that $\Delta^{1\pi}_{J\ell S}$ must fulfill  the set of $\ell-1$ sum rules (constraints)
\begin{align}
\int_{-\infff}^{L}\!\!\!\!\!\!\!\!\dd k^2 \frac{\Delta_{J\ell S}^{1\pi}(k^2)}{k^{2\lambda}}=0~,
\label{1pi.sr}
\end{align}
with $\lambda=2,3,\ldots,\ell$ and $\ell\geq 2$. These constraints are obtained straightforwardly by performing the expansion of Eq.~\eqref{1pi} in powers of $A$ and imposing that $T_{J\ell S}(A)\to A^{\ell}$ when $A\to 0$. 

Let us now consider again Eq.~\eqref{nn}. As $D_{J\ell S}(A)\to 1$ for $A\to 0$ then $T_{J\ell S}(A)\to N_{J\ell S}(A)$ in this limit. The expression for $N_{J\ell S}(A)$, Eq.~\eqref{nn}, is similar to Eq.~\eqref{1pi}. Indeed, they would be the same equation if $D_{J\ell S}(A)$ were replaced by 1 in Eq.~\eqref{nn} (and with $\Delta_{J\ell S}(k^2)$ evaluated at LO). As a result, $N_{J\ell S}(A)$, determined by implementing Eq.~\eqref{dis.rel.d} into Eq.~\eqref{nn}, does not vanish as $A^ \ell$ for $A\to 0$, because of the departure from 1 of $D_{J\ell S}(A)$ in an actual calculation.

It is convenient to proceed in a such a way that the right behavior of $T_{J\ell S}(A)$ around threshold is incorporated explicitly. For that purpose we consider the $N/D$ equation for $T_{J\ell S}(A)/A^\ell$, instead of $T_{J\ell S}(A)$. The quotient $T_{J\ell S}(A)/A^\ell$ has no  pole at $A=0$ because  $T_{J\ell S}(A){\to}A^\ell$ when $A\to 0$. Notice also that due to unitarity $T_{J\ell S}(A)/A^\ell\to 0$ for $A\to +\infff$. Then, according to the Sugawara and Kanazawa theorem \cite{barton,suga}, no subtractions are needed for the DR of $N_{J\ell S}(A)$ with $\ell>0$. EFT results do not always share the right high energy behavior so that subtractions will be certainly needed for a higher order calculation of $\Delta_{J\ell S}(A)$. At LO this is not the case because $\Delta_{J\ell S}^{1\pi}\to 1/A$ for $A\to \infff$. We then have ($\ell>0$)
\begin{align}
\label{for.2.t}
T_{J\ell S}(A)&=A^\ell \frac{N_{J\ell S}(A)}{D_{\jls}(A)}~,\\
\label{for.2.n}
N_{J\ell S}(A)&=\frac{1}{\pi}\int_{-\infff}^L\!\!\!\!\!\!\!\!\dd k^2 \frac{\Delta_{J\ell S}(k^2)D_{J\ell S}(k^2)}{k^{2\ell}(k^2-A)}~,\\
\label{for.2.d}
D_{J\ell S}(A)&=1-\frac{A}{\pi}\int_{0}^{\infff}\!\!\!\!\!\!\!\!\dd q^2 \frac{\rho(q^2)q^{2(\ell-1)}N_{J\ell S}(q^2)}{q^2-A}\nn\\
&=1+\frac{A}{\pi^2}\int_{-\infff}^L\!\!\!\!\!\!\!\!\dd k^2 \frac{\Delta_{J\ell S}(k^2)D_{J\ell S}(k^2)}{k^{2\ell}}  \nn\\
& \times \int_0^\infff\!\!\!\!\!\!\!\dd q^2 \frac{\rho(q^2)q^{2(\ell-1)}}{(q^2-A)(q^2-k^2)}~,
\end{align}
where the subtraction has been taken at threshold.

The previous equation for $D_{J\ell S}(A)$ is not satisfactory when $\ell\geq 2$ because the last integration on the right-hand side (r.h.s.) of Eq.~\eqref{for.2.d} is divergent. In this way, by applying the $N/D$ method to $T_{J\ell S}(A)/A^\ell$ we have changed the problem of the bad behavior of $T_{J\ell S}(A)$ around threshold into a high energy problem in the form of divergent integrals. To end up with a convergent DR for $D_{J\ell S}(A)$  in Eq.~\eqref{for.2.d} it is necessary that  $N_{J\ell S}(A)$ vanishes at least as
\begin{align}
N_{J\ell S}(A)\to 1/A^\ell ~~\text{for}~~A\to\infff~.
\label{n.infinity}
\end{align}
However, $N_{J\ell S}(A)$ vanishes only as $1/A$, independently of  $\ell$, as follows from Eq.~\eqref{for.2.n}. The set of constraints needed to satisfy the asymptotic behavior in Eq.~\eqref{n.infinity} can be deduced by performing in Eq.~\eqref{for.2.n} a high energy expansion of $N_{J\ell S}(A)$ in powers of $1/A$. It results in
\begin{align}
\int_{-\infff}^L\!\!\!\!\!\!\!\!\dd k^2 \frac{\Delta_{J\ell S}(k^2)D_{J\ell S}(k^2)}{k^{2\lambda}}=0~,
\label{n.sr}
\end{align}
with $\lambda=2,3,\ldots,\ell$ and $\ell\geq 2$. These sum rules generalize the ones fulfilled by $\Delta_{J\ell S}^{1\pi}(A)$ in Eq.~\eqref{1pi.sr}.\footnote{Equation~\eqref{n.infinity} is a consequence of Eq.~\eqref{for.2.t} because for $A\to +\infff$, due to unitarity, the ratio $T_{J\ell S}(A)/A^\ell$ tends to $1/A^{\ell+1/2}$ while $D_{J\ell S}(A)\to A^{1/2}$ (when only one subtraction is taken.) \label{foot.1}}

The usefulness of the $\ell-1$ restrictions in Eq.~\eqref{n.sr} can be well appreciated 
by rewriting  $N_{J\ell S}(A)$ in Eq.~\eqref{for.2.n} as 
\begin{align}
N_{J\ell S}(A)&=-\frac{1}{\pi}\sum_{m=0}^{\ell-2}\frac{1}{A^{m+1}}\int_{-\infff}^L 
\!\!\!\!\!\!\!\!\dd k^2 \frac{\Delta_{J\ell S}(k^2)D_{J\ell S}(k^2)}{k^{2(\ell-m)}}  \nn \\
& +\frac{1}{\pi A^{\ell -1}}\int_{-\infff}^L\!\!\!\!\!\!\!\!\dd k^2 \frac{\Delta_{J\ell S}(k^2)D_{J\ell S}(k^2)}{k^2(k^2-A)}~.
\end{align}
The last  term on the r.h.s. of the previous equation vanishes explicitly as $1/A^\ell$ for $A\to \infff$, while the terms in the sum on $m$ are zero once the constraints of Eq.~\eqref{n.sr} are fulfilled. In this way, inserting this expression for $N_{J\ell S}(A)$ 
in Eq.~\eqref{for.2.d} one has
\begin{align}
\label{int.d.3}
D_{J\ell S}(A)&=1+\frac{A}{\pi}\int_{-\infff}^L\!\!\!\!\!\!\!\!\dd k^2 \,\frac{\Delta_{J\ell S}(k^2)D_{J\ell S}(k^2)}{k^2}\, g(A,k^2) ~,
\end{align}
and a convergent DR integral equation for $D_{J\ell S}(A)$ results. 

It should be stressed that Eqs.~\eqref{for.2.t}, \eqref{for.2.n}, and \eqref{int.d.3} lead to the same equations as for the case of a $P$ wave amplitude, $\ell=1$, cf. Eqs.~\eqref{nd.def}, \eqref{dd_finalform} and \eqref{nn_finalform}.\footnote{It is equivalent to have the explicit factor  $A$ in $T_{\jls}$, Eq.~\eqref{for.2.t}, or included in the definition of $N_{\jls}$, Eq.~\eqref{nn_finalform}.} In the case of a $P$ wave, no constraints are needed because the right behavior  near threshold is  obtained straightforwardly. On the other hand, Eq.~\eqref{int.d.3} can be readily applied to $S$ wave by just adding the term proportional to $N_{J\ell S}(0)$ present in Eq.~\eqref{dd_finalform}. One subtraction should be taken in Eq.~\eqref{for.2.n} in order to transform it as Eq.~\eqref{nn_finalform} for $\ell=0$.

Now, let us address the way to solve the $N/D$ method, Eqs.~\eqref{for.2.t}, \eqref{for.2.n}, and \eqref{int.d.3}, in the presence of the constraints given in Eq.~\eqref{n.sr}.  It is well known \cite{spearman,barton} that the function $D_{J\ell S}(A)$ is determined modulo the addition of CDD poles \cite{cdd}. These are associated to specific dynamical features of the interaction that arise independently of the LHC discontinuity, $\Delta_{J\ell S}(A)$, and unitarity. Typically, the addition of CDD poles corresponds to the existence of pre-existing resonances or to Adler zeros \cite{nd,chew}. Both facts are indeed absent in the low-energy $NN$ scattering \cite{Stoks:1994wp}. We exploit this ambiguity in the function $D_{J\ell S}(A)$  and include $\ell-1$ CDD poles at {\it infinity}, so as to satisfy Eq.~\eqref{n.sr}:
\begin{align}
\label{d.con.cdd}
D_{J\ell S}&=1+\frac{A}{\pi}\int_{-\infff}^L\!\!\!\!\!\!\!\!\dd k^2 \frac{\Delta_{J\ell S}(k^2)D_{J\ell S}(k^2)}{k^2} g(A,k^2)\nn\\
& +\sum_{i=1}^{\ell-1}\frac{A}{B_i}\frac{\gamma_i}{A-B_i}~.
\end{align} 
The last term in the r.h.s. corresponds to adding the $\ell-1$ CDD poles. The factor $A/B_i$ in front of every CDD pole arises because the function $D_{J\ell S}(A)$ is normalized to 1 for $A=0$ and it has the residue $\gamma_i$ at $A=B_i$. The sum of the CDD poles gives rise to a rational fraction $Q_{\ell-1}/P_{\ell-1}$, where the subscript in $Q$ and $P$ indicate the degree of the polynomial in $A$.  Since the only relevant fact at low energies is the ratio $\gamma_i/B_i^2$ we take at the end the limit $B_i \to \infff$, with $\gamma_i/B_i^2$ not vanishing. The $\ell-1$ CDD poles are gathered at the same point $B$ and we write
\begin{align}
\sum_{i=1}^{\ell-1}\frac{A}{B_i}\frac{\gamma_i}{A-B_i}
\to \frac{A \sum_{n=0}^{\ell-2} c_n A^n}{(A-B)^{\ell-1}}~.
\label{cdd.f}
\end{align}
The coefficients $c_i$ are finally determined by requiring that the set of $\ell-1$ constraints in Eq.~\eqref{n.sr} are satisfied. The calculation is performed in terms of finite but large  $B$, and one has to check that the results are stable  by taking $B$ arbitrarily large. At the level of low-energy $NN$ scattering we have modified $D_{J\ell S}(A)$ by adding a polynomial of degree $\ell-1$ with fixed coefficients. 

We end with the following expressions:
\begin{align}
\label{for.2.t2}
&T_{J\ell S}(A)=A^\ell \frac{N_{J\ell S}(A)}{D_{\jls}(A)}~,\\
\label{for.2.n2}
&N_{J\ell S}(A)=\frac{1}{\pi}\int_{-\infff}^L\!\!\!\!\!\!\!\!\dd k^2 \frac{\Delta_{J\ell S}(k^2)D_{J\ell S}(k^2)}{k^{2\ell}(k^2-A)}~,\\
\label{for.2.d2}
& D_{J\ell S}(A)=1+\frac{A}{\pi}\int_{-\infff}^L\!\!\!\!\!\!\!\!\dd k^2 \frac{\Delta_{J\ell S}(k^2)D_{J\ell S}(k^2)}{k^2} g(A,k^2)\nn \\
& +\frac{A \sum_{n=0}^{\ell-2} c_n A^n}{(A-B)^{\ell-1}}~,
\end{align}
and the constraints 
\begin{align}
&\int_{-\infff}^L\!\!\!\!\!\!\!\!\dd k^2 \frac{\Delta_{J\ell S}(k^2)D_{J\ell S}(k^2)}{k^{2\lambda}}=0~,\ \lambda=2,3,\ldots,\ell~,~\ell\geq 2~.
\label{const.sueltas}
\end{align}
The previous formalism is also meaningful for the case in which $\Delta_{J\ell S}(A)\to C$, with $C$ a constant, as $A\to \infff$. We do not discuss in this work its extension to the case when $\Delta_{J\ell S}(A)$ diverges for $A\to \infff$, as we are interested now in LO $NN$ scattering. This generalization of our formalism will be discussed when considering higher orders in the chiral expansion of $\Delta_{J\ell S}(A)$ \cite{future}, which include the important TPE contributions \cite{Rentmeester:1999vw,Rentmeester:2003mf}.

Summarizing the results of this section, we have presented a general approach based on the $N/D$ method to construct $NN$ scattering partial wave amplitudes. For $\ell=0,1$ Eqs.~\eqref{dd_finalform} and \eqref{nn_finalform} are employed, with $T_{\jls}$ given by Eq.~\eqref{nd.def}. For $\ell \ge 2$, one has Eqs.~\eqref{for.2.t2}--\eqref{for.2.d2}, that must be solved in the presence of the constraints given in Eq.~\eqref{const.sueltas}. The solution of this integral equation subjected to the constraints is discussed in the Appendix.

\section{Results}
\label{sec:results}

\begin{figure}[h]
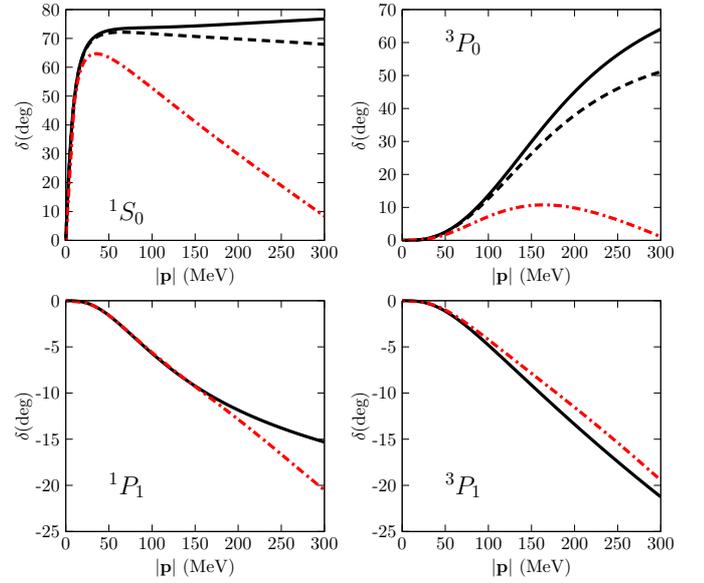
\centering
\begin{tabular}{cc}
\includegraphics[height=3.75cm,keepaspectratio]{figurasNN_18.eps} &
\includegraphics[height=3.75cm,keepaspectratio]{figurasNN_17.eps} \\
\includegraphics[height=3.75cm,keepaspectratio]{figurasNN_16.eps} &
\includegraphics[height=3.75cm,keepaspectratio]{figurasNN_15.eps}
\end{tabular}
\caption{Comparison of the $^1 S_0$, $^3 P_0$, $^1 P_1$, and $^3 P_1$ phase shifts with the Nijmegen PWA data. The solid (black) lines represent the results of this work, while the dash-dotted (red) ones represent the Nijmegen PWA \cite{Stoks:1994wp}. For the $^1 S_0$ and $^3 P_0$, the dashed (black) lines correspond to the relativistic version of this work, see the text for details.\label{elas.nocdd.fig}}
\end{figure}

In this section we present our results for the phase shifts, $\delta$, of the elastic partial waves with $\ell\leq 5$. We compare them with the Nijmegen partial wave analysis (PWA) \cite{Stoks:1994wp} in Figs.~\ref{elas.nocdd.fig}--\ref{elas2.cdd.fig}. Our results are represented  by the solid (black) lines, and the Nijmegen data by the dash-dotted (red) lines. We show the results up to $\lvert \vp \rvert = 300\ \textrm{MeV}$. Notice that the pion production threshold opens at $\lvert \vp \rvert \simeq 360\ \textrm{MeV}$ and the three-momentum is no longer small, $|\vp|\simeq \sqrt{m m_\pi}\gg m_\pi$.

In Fig.~\ref{elas.nocdd.fig} we show the lowest elastic waves, namely, $^1 S_0$, $^1 P_1$, $^3 P_1$, and $^3 P_0$, whose amplitudes do not contain CDD poles because $\ell<2$. The agreement in the $^1 P_1$ and $^3 P_1$ partial waves is quite satisfactory. For the $^1 S_0$ it is known that a higher order chiral counterterm is needed to reproduce the large effective range and thus improve the agreement with the data \cite{nogga}. In the case of the $^3 P_0$ wave, large corrections stem from TPE. Since this is a LO calculation, none of these corrections is included. However, an important point should be stressed: in this paper a regulator (cut-off) independent and unitary description of the $NN$ interaction with the right analytical properties is reached. The agreement with the data can be improved by including higher orders in the LHC.

For the $^1 S_0$ and $^3 P_0$ waves we have also tried with a \emph{relativistic} calculation of the function $g(A,k^2)$, Eq.~\eqref{g.def}, since these are the waves for which the discrepancies with the data are larger. In this approach, the $\rho(q^2)$ function is replaced by its relativistic counterpart in the $S$-matrix and in the integrals where it is involved,
\begin{align*}
\rho(q^2) = \frac{\sqrt{q^2} m}{4\pi} \to \rho(q^2) = \frac{\sqrt{q^2} m}{4\pi} \frac{m}{\sqrt{q^2 + m^2}}~.
\end{align*}
The results obtained are represented by the dashed (black) lines in Fig.~\ref{elas.nocdd.fig}. Though the corrections are in the right direction, the discrepancies with the data are still large. As expected, relativistic corrections are small in the energy range shown, though noticeable for the $^3P_0$ partial wave for $|\vp|\gtrsim 200$~MeV.

\begin{figure}[ht]\centering
\begin{tabular}{cc}
\includegraphics[height=3.75cm,keepaspectratio]{figurasNN_13.eps} &
\includegraphics[height=3.75cm,keepaspectratio]{figurasNN_11.eps} \\
\includegraphics[height=3.75cm,keepaspectratio]{figurasNN_10.eps} &
\includegraphics[height=3.75cm,keepaspectratio]{figurasNN_9.eps}
\end{tabular}
\caption{Comparison of the $^1 D_2$, $^3 D_2$, $^1 F_3$ and $^3 F_3$ phase shifts. For the notation, see Fig.~\ref{elas.nocdd.fig}.\label{elas1.cdd.fig}}
\end{figure}

\begin{figure}[ht]\centering
\begin{tabular}{cc}
\includegraphics[height=3.75cm,keepaspectratio]{figurasNN_19.eps} &
\includegraphics[height=3.75cm,keepaspectratio]{figurasNN_20.eps} \\
\includegraphics[height=3.75cm,keepaspectratio]{figurasNN_21.eps} &
\includegraphics[height=3.75cm,keepaspectratio]{figurasNN_22.eps}
\end{tabular}
\caption{Comparison of the $^1 G_4$, $^3 G_4$, $^1 H_5$ and $^3 H_5$ phase shifts. For the notation, see Fig.~\ref{elas.nocdd.fig}.\label{elas2.cdd.fig}}
\end{figure}

For higher waves one needs to include $\ell-1$  CDD poles in order to fulfill the constraints in Eq.~\eqref{const.sueltas}. Thus, it is guaranteed that the partial waves have the right behavior at threshold, vanishing as $|\vp|^{2\ell}$. Our results are shown in Figs.~\ref{elas1.cdd.fig}~and~\ref{elas2.cdd.fig}, and a good agreement with the data is achieved, except for the $^1 D_2$ wave. Our curves are quite similar compared with the LO results of Ref.~\cite{nogga}. This reference offers an approach with cut-off independent results with the $NN$ potential ($V_{NN}$) given by OPE. The largest discrepancy concerns to the $^3 P_0$ partial wave where in Ref.~\cite{nogga} a counterterm is promoted from higher orders so as to achieve cut-off stable results for $\Lambda\to\infff$ due to the attractive $1/r^3$ tensor force in OPE. As a result, the agreement with data is much improved. The main difference between our approach and that of Ref.~\cite{nogga} is the treatment of the LHC. Namely, for the $^3P_0$ wave the iteration of the $NN$ potential is responsible for the need of this extra counterterm. The first iteration $V_{NN} G V_{NN}$ (with $G$ the unitary two-nucleon reducible loop function and $V_{NN}$ the nucleon potential) is a new source of LHC discontinuity \cite{pdbcollins}  containing contributions from TPE and iterated OPE. The real part stemming from the former is divergent. Within our approach the sources of LHC discontinuity from  $V_{NN} G V_{NN}$ are NLO according with the standard chiral counting. At that order new subtractions are required \cite{future} which will mimic the role of the extra counterterm taken in Ref.~\cite{nogga}. While our method is based on the calculation of $\Delta_{J\ell S}(A)$ perturbatively along the LHC, the application of a Lippmann-Schwinger equation with the chiral $NN$ potentials is based on the perturbative calculation of the latter \cite{w1,w2,w3}. In both cases the diagrams required for the calculation of $\Delta_{J\ell S}$ and $V_{NN}$ are two-nucleon irreducible, which justifies its perturbative treatment \cite{long,let1,w1,w2,w3}. In both cases as well the RHC is exactly resummed, as required because of the enhanced two-nucleon reducible diagrams. This resummation is performed in terms of the interaction kernel, $\Delta_{J\ell S}$ or $V_{NN}$, depending on the approach. The $N/D$ method respects the LHC discontinuity so that $\Delta_{J\ell S}$ is the same as in the final partial wave amplitude. For a Lippmann-Schwinger equation this is not the case as new sources of imaginary parts along the LHC result from the iteration of $V_{NN}$ \cite{pdbcollins}. It is also worth stressing that our approach based on the $N/D$ method is a dispersive one offering results that by construction are cut-off independent, while this is still an issue in the application of the Lippmann-Schwinger (or Schr\"odinger) equation to $NN$ scattering with $V_{NN}$ calculated from ChPT \cite{nogga,pav11,epege,epemeiss06}.

For the $^1 S_0$ and $^1 D_2$ partial waves, for which we do not have good agreement at LO with data \cite{Stoks:1994wp}, our results are indeed very similar to those of Ref.~\cite{nogga}, too. In the case of the $^1P_1$ partial wave our phase shifts run closer to data at low energies than those of Ref.~\cite{nogga}. 

In order to show the independence of our results with the value of $B$, the position of the CDD poles, once this value is large enough, we show in Fig.~\ref{elas2.cdd.fig} for the $^3 G_4$ partial wave different lines corresponding to $B=10^n m_\pi^2$ for $n=2,3,4,5\text{, and}\ 6$. A narrow band is obtained despite the large variation in the values of $B$ considered.

\section{Conclusions}
\label{sec:conclusions}

We have applied the $N/D$ method to $NN$ scattering from chiral perturbation theory. In this method the two cuts present in a $NN$ partial wave, the right-hand cut and left-hand cut, are separated in two functions, $D_{J\ell S}(A)$ and $N_{J\ell S}(A)$, with $A$ the center-of-mass three-momentum squared. While $D_{J\ell S}(A)$ has only right-hand cut the function $N_{J\ell S}(A)$ has only left-hand cut. The $NN$ partial waves, $T_{J\ell S}=N_{J\ell S}/D_{J\ell S}$ ($\ell=0,~1$) and $A^\ell N_{J\ell S}/D_{J\ell S}$ ($\ell\geq 2$), are determined in terms of their discontinuity along the left-hand cut due to multipion exchanges, $\Delta_{J\ell S}(A)$. At leading order, considered in this work,  only OPE contributes. For $D$- and higher partial waves, with orbital angular momentum $\ell\geq 2$, one has to impose the proper behavior of a partial wave near threshold, such that it vanishes as $A^\ell$ for $A\to 0$. This gives rise to $\ell-1$ constraints in the form of sum rules involving the functions $\Delta_{J\ell S}$ and $D_{J\ell S}$. Since the function $D_{J\ell S}(A)$ is determined modulo the addition of Castillejo-Dalitz-Dyson poles (that correspond to  zeros of the $NN$ partial waves along the real axis) we have added $\ell-1$ of such poles at infinity in $D_{J\ell S}$   for $\ell\geq 2$. By sending such poles to infinity no zero at finite energies is included for any $NN$ partial wave. In addition, the residues of these poles in $D_{J\ell S}$ are fixed once the sum rules are satisfied, so that no new parameters are included. At low energies the Castillejo-Dalitz-Dyson poles behaves like adding a polynomial of degree $\ell-2$ to $D_{J\ell S}$. 

  The resulting $NN$ partial waves do not contain any regulator. A subtraction constant is required for the $^1S_0$ partial-wave that is fixed by reproducing the experimental scattering length. Our results are very close to those of Nogga, Timmermans, and van Kolck \cite{nogga} that provide cut-off independent $NN$ partial waves with OPE as potential. The only noticeable difference concerns the $^3P_0$ partial wave for which Ref.~\cite{nogga} achieves cut-off stable results by promoting a higher-order counterterm to leading order due to the attractive $1/r^3$ tensor force in OPE. In our approach there is no special treatment for the $^3P_0$  partial wave compared to others because of the perturbative treatment of the discontinuity across the LHC. Our results are a prediction for the $^3P_0$ phase shifts at leading order. For the $^1 P_1$ partial wave our phase shifts run closer to data than those of \cite{nogga}. 

  This method is ready to be extended to higher orders and it would be of great interest to apply it including  TPE contributions to $\Delta_{J\ell S}$. TPE is needed for an accurate description of the $NN$ data \cite{Rentmeester:1999vw,Rentmeester:2003mf}. Its extension to coupled channels with $\ell=J\pm 1$ is being studied \cite{albaprepare}.

\begin{acknowledgments}
 J.A.O. would like to acknowledge useful discussions with Ubirajara van Kolck. We also acknowledge Manuel Pav\'on Valderrama for discussions and a critical reading of the manuscript. This work is partially funded by grants no. MEC  FPA2010-17806 and the Fundaci\'on S\'eneca 11871/PI/09. M.A. acknowledges the Fundaci\'on S\'eneca grant no. 13310/FPI/09. We also thank the financial support from the BMBF grant no. 06BN411, the EU-Research Infrastructure Integrating Activity ``Study of Strongly Interacting Matter" (HadronPhysics2, grant no. 227431) under the Seventh Framework Program of EU and the Consolider-Ingenio 2010 Programme CPAN (CSD2007-00042).

\end{acknowledgments}

\appendix
\titleformat{\section}[hang]
{}
{\filcenter}
{5pt}
{\filcenter\bfseries\MakeUppercase  APPENDIX: \bfseries\MakeUppercase}


\section{Solving the integral equation for $D(k^2)$}
\label{app.sol.d}
In this appendix, we focus on the solution of the integral equation \eqref{for.2.d2} subject to the constraints \eqref{const.sueltas}. For simplifying the discussion we drop the subscripts $J\ell S$. The integral equation Eq.~\eqref{for.2.d2} can be written in a compact way as
\begin{align}\label{int.d.may}
D(A) & = 1 + \frac{A}{\pi}\int_{-\infff}^{L}\!\!\!\!\!\!\!\!\dd k^2 \frac{\Delta(k^2)D(k^2)}{k^2} g(A,k^2) + h(A)~\text{,}
\end{align}
where $g(A,k^2)$ is defined in Eq.~\eqref{g.def} and
\begin{align}
h(A)     & = \frac{A}{(A-B)^{\ell - 1}} \sum_{i=0}^{\ell - 2} c_i A^i~\text{.}
\end{align}
Let us introduce the function $d(A)$ as $D(A) = d(A) + h(A)$, that is, $d(A)$ is the piece of $D(A)$ that does not contain the CDD poles. As a first step, we write the coefficients $c_i$ in terms of the $d(A)$ function. This is done by writing the sum rule constraints Eq.~\eqref{const.sueltas} in terms of $d(A)$, giving rise to
\begin{align}
I_i & = \int_{-\infff}^{L}\!\!\!\!\!\!\!\!\dd k^2 \frac{\Delta(k^2)d(k^2)}{k^4 k^{2i}} = - \int_{-\infff}^{L}\!\!\!\!\!\!\!\!\dd k^2 \frac{\Delta(k^2)h(k^2)}{k^4 k^{2i}} \nn\\
& = \sum_{j=0}^{\ell - 2} c_j m_{ij}~\text{,}\\
& m_{ij} \equiv -\int_{-\infff}^{L}\!\!\!\!\!\!\!\!\dd k^2 \frac{\Delta(k^2) k^{2j}}{k^2 k^{2i}(k^2-B)^{\ell-1}}~\text{,}
\end{align}
where we have shifted $\lambda = i + 2$, so that $i$ runs from $0$ to $\ell - 2$. Note that the integrals $m_{ij}$ can be calculated directly for a given $\Delta(A)$ in terms of $B$ because the unknown function $d(A)$ does not appear in their calculation. Thus, 
\begin{align}\label{coeff.ci}
c_i & = \sum_{j=0}^{\ell - 2} \left. m^{-1} \right\rvert_{ij} I_j~\text{,}
\end{align}
being $m^{-1}$ the inverse of the matrix $\lvert\lvert m_{ij} \rvert\rvert$. Next, we rewrite the integral equation Eq.~\eqref{int.d.may} in terms of $d(k^2)$, and insert Eq.~\eqref{coeff.ci} for the coefficients $c_i$. It results 
\begin{align}
d(A)  = & 1 + \frac{A}{\pi}\int_{-\infff}^{L}\!\!\!\!\!\!\!\!\dd k^2 \frac{\Delta(k^2)d(k^2)}{k^2} g(A,k^2) \nn \\
& + \frac{A}{\pi} \int_{-\infff}^{L}\!\!\!\!\!\!\!\!\dd k^2 \frac{\Delta(k^2) g(A,k^2)}{(k^2 - B)^{\ell-1}} \sum_{i,j=0}^{\ell - 2}\!\! k^{2i}\!\!\left.m^{-1}\right\rvert_{ij}\!\!\! \nn \\ & \times \int_{-\infff}^{L}\!\!\!\!\!\!\!\!\dd q^2 \frac{d(q^2) \Delta(q^2)}{q^4 q^{2j}}~\text{.}
\end{align}
Now, by interchanging the dummy integration variables $k^2$ and $q^2$, we can finally write
\begin{align}
d(A) & = 1+ \frac{A}{\pi}\int_{-\infff}^{L}\!\!\!\!\!\!\!\!\dd k^2 \frac{d(k^2) \Delta(k^2)}{k^2} \left( g(A,k^2) + \bar{g}(A,k^2) \right)~\text{,} \\
\bar{g}(A,k^2) & = \sum_{i,j=0}^{\ell-2} \frac{1}{k^{2j}} m^{-1}_{ij}\int_{-\infff}^{L}\!\!\!\!\!\!\!\!\dd q^2 \frac{\Delta(q^2) g(A,q^2) }{(q^2 - B)^{\ell - 1}}~\text{.} \nn
\end{align}
We have now an integral equation for the $d(k^2)$ function that depends on known functions. It can be written in a compact way as
\begin{align}
d(A) = 1 + \int_{-\infff}^{L}\!\!\!\!\!\!\!\!\dd k^2 \widetilde{f}(A,k^2) d(k^2)~\text{.}
\end{align}
It is convenient to perform a change of integration variable so that one ends with a finite integration domain, e.g. with $x=1/k^2$. In this way
\begin{align}
d(A) = 1 + \int_{x_1}^{x_2}\!\!\!\!\!\!\!\!\, dx \,f(A,x)\, d(k^2(x))~\text{.}
\end{align}
 
This is an inhomogeneous Fredholm integral equation. We solve it numerically  by discretizing the integral on it,
\begin{align}
d(A_i) = 1 + \sum_{j}  \,f(A_i,x_j) \omega(x_j) d(k^2(x_j)) ~\text{,}
\end{align}
where the $\omega(x)$ function is the weighting function taken for the integration. By calling $d(k^2(x_i)) \equiv d_i$, $f(A_i,x_j)\omega(x_j) \equiv \eta_{ij}$, this equation can be recast as
\begin{align}
\sum_{j} (\delta_{ij} - \eta_{ij}) d_j = 1~\text{,}
\end{align}
which is a linear equation, that can be solved by standard methods, giving the desired function $d(A)$. To obtain $D(A)$, the function $h(A)$ must be added, but this is also a direct task, since the $c_i$ coefficients can be calculated once $d(A)$ is known, Eq.~\eqref{coeff.ci}. Of course, if no constraints must be satisfied, as it is the case for $S$ and $P$ waves, the same formalism with the corresponding simplifications should be used.



\begin{thebibliography}{99}
\bibitem{future} In this case one is forced to take more subtractions in the DR for $D_{J\ell S}(A)$, Eq.~\eqref{for.2.d}. Depending on the degree of 
divergence of $\Delta_{J\ell S}(A)$ they can also be necessary in the DR for $N_{J\ell S}(A)$, Eq.~\eqref{for.2.n}. 
 J.~A.~Oller {\it et al.}, work in progress.
\end{thebibliography}
\end{document}